\documentclass[useAMS,usenatbib]{mn2e}

\usepackage{calc}
\usepackage{fancyhdr}
\usepackage[dvips]{graphicx}
\usepackage[normalem]{ulem}
\usepackage{url}

\newcommand{\eqncomma}{\mbox{~,}}

\newcommand{\eqnperiod}{\mbox{~.}}

\newcommand{\ie}{\emph{i.e.}}
\newcommand{\ulem}[1]{\emph{\uline{#1}}}

\title[A Moving Group $\sim 50$ kpc from the Sun]
{Statistical Properties of Blue Horizontal Branch Stars in the Spheroid: Detection of a Moving Group $\sim 50$ kpc from the Sun}

\author[Harrigan et al.]
{Matthew J. Harrigan,$^{1}$\thanks{E-mail:harrim6@rpi.edu,newbeh@rpi.edu}
Heidi Jo Newberg,$^{1}$
Lee A. Newberg,$^{2,3}$
Brian Yanny,$^{4}$
\newauthor
Timothy C. Beers,$^{5}$
Young Sun Lee$^{5}$ and
Paola Re Fiorentin$^{6,7}$\\
$^{1}$Dept. of Physics, Applied Physics and Astronomy, Rensselaer
Polytechnic Institute Troy, NY 12180\\
$^{2}$Dept. of Computer Sciences, Rensselaer Polytechnic Institute, Troy, NY 12180\\
$^{3}$Wadsworth Center, New York State Department of Health, Albany, NY, 12201\\
$^{4}$Fermi National Accelerator Laboratory, P.O. Box 500, Batavia, IL 60510\\
$^{5}$Dept. of Physics \& Astronomy, CSCE: Center for the Study of Cosmic Evolution and \\
JINA: Joint Institute for Nuclear Astrophysics, Michigan State University, E. Lansing, MI 48824\\
$^{6}$Department of Physics, University of Ljubljana, Jadranska 19, 1000 Ljubljana, Slovenia\\
$^{7}$Max-Planck-Institut f\"{u}r Astronomie, K\"{o}nigstuhl 17, D-69117 Heidelberg, Germany}

\begin{document}

\date{Accepted for publication in MNRAS 2010}

\pagerange{\pageref{firstpage}--\pageref{lastpage}} \pubyear{2009}

\maketitle

\label{firstpage}

\begin{abstract}

A new moving group comprising at least four Blue Horizontal Branch (BHB) stars is identified
at $(l,b)=(65^\circ, 48^\circ$).
The horizontal branch at $g_0=18.9$ magnitude implies a distance of 50 kpc from the Sun.
The heliocentric radial velocity is $\langle V_r\rangle=-157\pm4$ km s$^{-1}$, corresponding to
$V_{\mbox{\it gsr}}=-10$ km s$^{-1}$; the dispersion in line-of-sight velocity is consistent with
the instrumental errors for these stars.
The mean metallicity of the moving group is [Fe/H]$\sim -2.4$,
which is significantly more metal poor than the stellar spheroid.  
We estimate that the BHB stars in the outer halo have a mean
metallicity of [Fe/H]=$-2.0$, with a wide scatter and a distribution that does
not change much as a function of distance from the Sun.  We explore the systematics
of SDSS DR7 surface gravity metallicity determinations for faint BHB stars, and present a technique
for estimating the significance of clumps discovered in multidimensional
data.  This moving group cannot be distinguished in density, and highlights 
the need to collect many more spectra of Galactic stars to unravel the 
merger history of the Galaxy.
\end{abstract}

\begin{keywords}
Galaxy: structure --- Galaxy: halo --- methods: statistical.
\end{keywords}

\section{Introduction}

In recent years, stellar photometry primarily from the Sloan Digital Sky Survey (SDSS)
has allowed astronomers to discover previously unknown spatial substructure 
in the Galactic spheroid.  Other surveys such as 2MASS and QUEST have also been
influential.  Highlights include the discovery of new tidal
streams \citep{2002ApJ...569..245N,2003AJ....126.2385O,2006ApJ...639L..17G,
2006ApJ...643L..17G,2006ApJ...645L..37G,2006ApJ...642L.137B,2009ApJ...700L..61N}; dwarf 
galaxies, star clusters, and transitional objects
\citep{2005AJ....129.2692W,2005ApJ...626L..85W,2006ApJ...643L.103Z,2006ApJ...650L..41Z,
2006ApJ...647L.111B,2007ApJ...654..897B,2007ApJ...656L..13I,2009ApJ...696.2179K,2009MNRAS.397.1748B}; and previously 
unidentified spatial structure of unknown or controversial
identity \citep{2001ApJ...554L..33V,2003ApJ...588..824Y,2004ApJ...615..738M,
2005ASPC..338..210N,2007ApJ...657L..89B,2008ApJ...673..864J,anVOD}.  

The discovery of spatial substructure immediately
suggests the need for spectroscopic follow-up to determine the orbital
characteristics of tidal debris, the masses of bound structures, and the
character of structures of unknown identity.  It has long been known that
a spheroid that is formed through accretion will retain the record of its
merger history much longer in the velocities of its component stars than it
will in their spatial distribution \citep{2003MNRAS.339..834H}.  For this
reason early searches concentrated on velocity to find ``moving groups"
in the spheroid \citep{1996ApJ...459L..73M} rather than on the search 
for spatial substructure.  

Groups of spheroid stars that have similar
spatial positions and velocities (moving groups) are generally equated with
tidal disruption of star clusters or dwarf galaxies, unlike local disc
``moving groups" that were once hypothesized to be the result of
disruption of star clusters \citep{bok,eggen} but are now thought to
be the result of orbital resonances \citep{2008A&A...483..453F}.

Recently, searches for spheroid velocity substructure in SDSS data have concentrated
on local (up to 17.5 kpc from the Sun) metal-poor main sequence stars 
\citep{klement,smith2009,schlaufman}, finding more than a dozen groups of stars
with coherent velocities.  \citet{schlaufman} estimate that there are $10^3$
cold substructures in the Milky Way's stellar halo.  Blue horizontal branch (BHB)
stars are particularly important for studies of spheroid substruscture
(e.g. Clewey \& Kinman 2006) because they can be seen to large
distances in the SDSS and because a large fraction of the SDSS stellar
spectra are BHBs.  In the future we need
more complete spectroscopic and proper motion surveys 
designed to discover velocity substructure in the Milky Way's spheroid.

In this paper we present evidence for a tidal moving group of BHB stars, discovered in the
SDSS and Sloan Extension for Galactic Understanding and Exploration (SEGUE; Yanny et al. 2009) 
spectroscopic survey.
Additional evidence that the stars are part of a coherent structure
comes from the unusually low metallicity of the stars in the moving group.
We are unable to isolate this moving group in density
substructure, highlighting the power of velocity
information to identify low density contrast substructure in the
spheroid.

Historically many of the spheroid substructures have been discovered by
eye in incomplete datasets, rather than by a mechanical analysis of data
with a well-understood background population.  As the size of the substructures
decreases, it becomes more critical to determine whether the observed structure
could be a random fluctuation of the background.  In this paper, we present
a method (Appendix A) for estimating the probability that a random fluctuation
could produce an observed ``lump."

As we work towards the fainter limits of the SDSS data, it becomes more important
to understand how stellar parameters derived from spectra are affected by a low 
signal-to-noise (S/N) ratio.  The stars in the newly detected halo substructure have
S/N ratio between 7 and 10, as measured from the ratio of fluctuations to continuum
on the blue side of the spectrum.  In the past we have successfully used spectra with S/N
as low as 5 to measure radial velocities of F turnoff stars in the Virgo
Overdensity \citep{netal07}.  In this paper, we show that although higher S/N is 
preferable, some information about metallicity and surface gravity can be gained
for BHB stars with S/N ratios less than 10.

Since the strongest evidence that these BHBs form a coherent group is that
they have unusually low metallicity, even for the spheroid, we also explore
the SEGUE metallicity determinations for BHB stars in the outer spheroid.
A great variety of previous authors have measured a large metallicity
dispersion for spheroid stars, with average metallicities somewhere in the
$-1.5 < \rm [Fe/H] < -1.7$ range \citep{1987ARA&A..25..603F,gwk89}.  These 
investigations analysed globular clusters 
\citep{1978ApJ...225..357S,1985ApJ...293..424Z}, 
RR Lyraes \citep{1985ApJ...289..310S,1991ApJ...367..528S}, 
K giants \citep{2003AJ....125.2502M}, and
dwarfs \citep{1990AJ.....99..201C}. \citet{1986ApJS...61..667N} found an
average [Fe/H] of $-1.67$ for globular clusters in the outer halo, and
an average metallicity of $-1.89$ for field stars in the outer halo, again with
a large metallicity dispersion that does not depend on distance.
The Besan\c{c}on model of the Milky Way adopted an average [Fe/H] of $-1.7$, with
$\sigma=0.25$, for the stellar halo \citep{2000A&A...359..103R}.
Recently, \citet{carollo,carollo2} analysed the full space motions of $>10,000$ stars
within 4 kpc from the Sun, and found evidence that the ones that are
kinematic members of the outer halo have an average metallicity of 
[Fe/H]=$-2.2$.

Our analysis of the SDSS BHB stars shows that the metallicity of BHB
stars in the Galactic spheroid does not change from $g_0=14.5$ (6 kpc
from the Sun) to $g_0=19.15$ (55 kpc from the Sun) at high Galactic
latitudes.  The mean measured metallicity of these stars is [Fe/H]=$-1.9$,
but analyses of globular clusters in the sample show that there could
be a systematic shift in the BHB measured metallicities of a few
tenths of a magnitude, and in fact [Fe/H]=$-2.0$ is our best guess for
the proper calibrated value.

\section {Observations and Data}

\begin{figure*}
%\plotone{fig1test.ps}
\noindent
%\centerline{\includegraphics[angle=-90,width=.95\textwidth]{fig1test.ps}}
\centerline{\includegraphics[angle=-90,width=.95\textwidth]{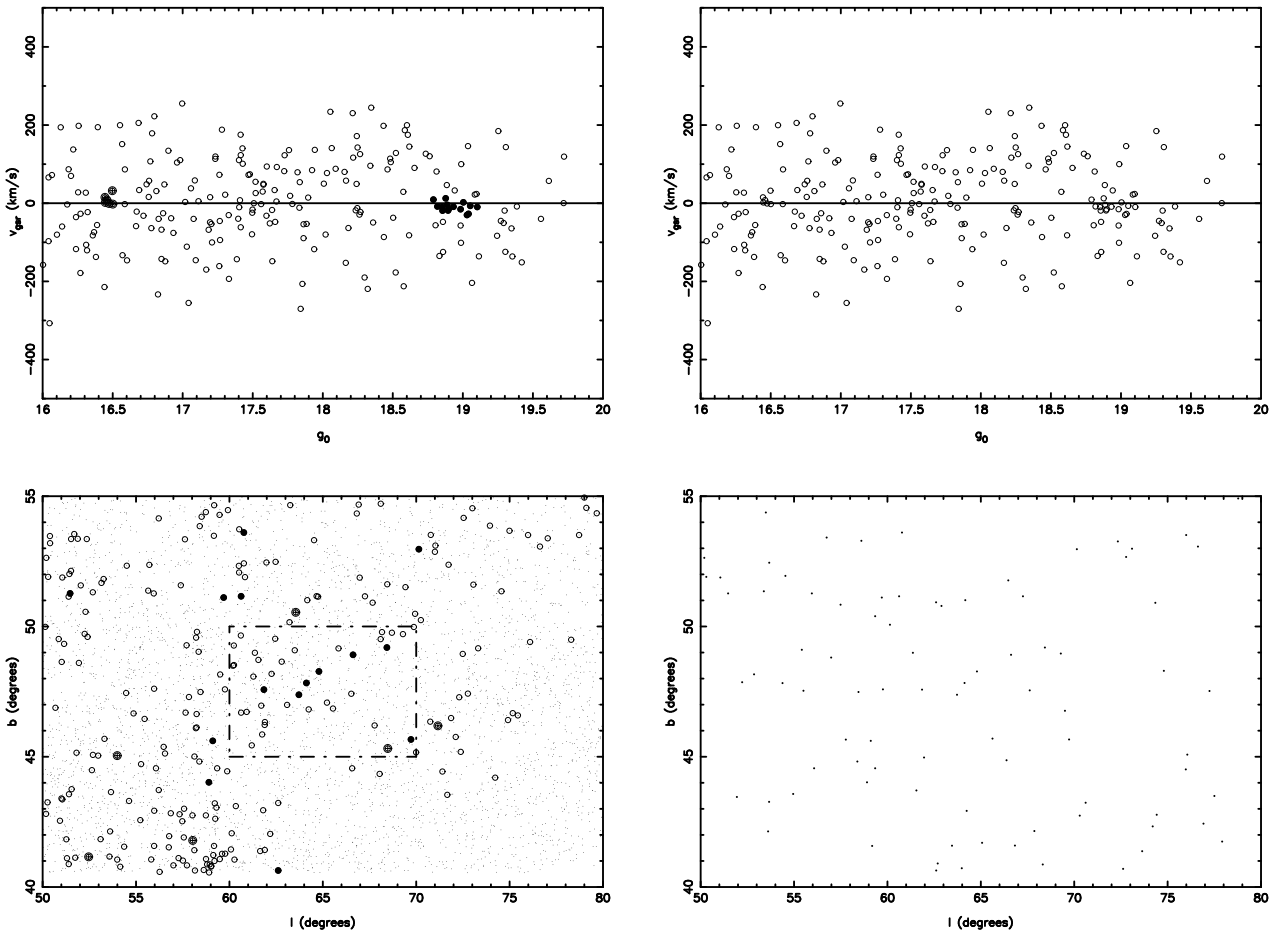}}
%\vspace{11pc}
\caption{Identification of a moving group.  The upper left and right plots show blue 
horizontal branch (BHB) stars with spectroscopic data in the region 
$50^\circ<l<80^\circ$ and $40^\circ<b<55^\circ$, where the velocity 
is the line of sight velocity with respect to the Galactic standard of rest.  The 
data in the two plots are the same, with the upper left plot marking two 
overdensities noticed on the upper right plot.  
Open circles in the lower left plot show the sky positions of the same BHB candidates
with spectroscopic data in SDSS DR6.  Filled circles correspond 
to stars that are in the co-moving stellar association, with the same stars plotted 
in the upper left and lower left plots.  These stars have $g_0$ 
magnitudes of $18.65 < g_0 < 19.15$ and line of sight velocities, with respect to 
the Galactic standard of rest, of $-35 < V_{\mbox{\it gsr}} < 15$ km s$^{-1}$.  The 
crossed circles show the positions of stars in a density at $g_0=16.5$ found on the upper left 
plot.  In the lower left figure, a higher fraction of the BHB stars have spectra near 
the globular cluster M13, at $(l,b)=(59^\circ,41^\circ)$.  The box indicates the area of 
the newly discovered over-density.  
The smaller black dots in the lower left plot indicate the 
positions of photometrically selected BHB stars in Galactic coordinates.  
The lower right plot shows the positions of the photometric 
data from the lower left plot that has the $g_0$ magnitude range of the 
moving group, $18.65 < g_0 < 19.15$.  Note that the moving group is not detected from the
photometric data alone.
\label{ident}} 
\end{figure*}

In order to search the stellar halo for co-moving groups of stars, we selected BHB
stars from the sixth data release (DR6; Adelman-McCarthy et al. 2008) of the SDSS.  
This release contains both Legacy Survey and SEGUE photometric data from 9583 square 
degrees of sky.  More technical information for the SDSS survey can be found 
in \citet{2000AJ....120.1579Y,1996AJ....111.1748F,1998AJ....116.3040G,2002AJ....123.2121S,
2002AJ....123..485S,2003AJ....126.2081A,2003AJ....125.1559P,
2004AN....325..583I,2006AJ....131.2332G,2006AN....327..821T}.

We selected 8753 spectra with the colors of A stars from the SDSS DR6 Star database using the 
color cuts $-0.3 < (g-r)_0 < 0.0$ and $0.8 < (u-g)_0 < 1.5$, within the magnitude 
range $15 < g_0 < 23$, where the $0$ subscript indicates that the magnitudes have been 
corrected for reddening and extinction using the \citet{1998ApJ...500..525S} reddening map.  
Within this color box, the stars that are bluer in $(u-g)_0$ tend to be high surface
gravity blue straggler (BS) or main sequence stars, and those that are redder in 
$(u-g)_0$ tend to be low surface gravity blue horizontal branch (BHB) stars.  Using 
$(u-g)_0$, $(g-r)_0$ color selection parameters as described in Figure 10 
of \citet{ynetal00}, we divided the sample into $4630$ candidate BHB 
stars and $4123$ candidate BS stars with spectra.  

We separated the stars in surface
gravity using a photometric technique, because in 2007 when we were selecting the stars,
we were concerned that the S/N of the spectra would degrade more quickly than the 
S/N of the photometry at the faint end of our sample.  The S/N of the spectroscopy increases
over the whole magnitude range from $15<g_0<20.5$, while the photometric accuracy degrades
only for $u>19$.  The photometric degradation affects halo BHB stars fainter than about $g_0=18$.
With more recent reduction software (see below), we have found similar sensitivity to surface gravity
in photometric and spectroscopic indicators for faint SDSS spectra, but the spectroscopic
separation is much better than photometric separation for bright, high S/N spectra.
The photometric selection 
technique does not introduce gradients in the selection effect with apparent magnitude,
and can be applied identically for stars with and without measured spectra.  

At bright magnitudes, the spectral S/N is high enough that the DR6 surface 
gravities are reliable.  The SDSS spectra are R=1800, and the S/N of A star spectra varies
from 50 at $g_0=16$ to 7 at $g_0=19$.  We tested the efficiency and completeness
of the photometric selection technique using 3996 SDSS spectra of stars with
$16<g_0<17.5$ and colors of A-type stars.  Of the 1948 that had surface gravities
consistent with BHB stars, 84\% were classified as BHBs by the photometric technique.
Of the 2048 spectra with high surface gravity (BS stars), 35\% were classified as
BHBs in the photometric technique.  These numbers for completeness and contamination
are similar to estimates that we made from examination of Figure 13 from
\citet{ynetal00}, and apply only to stars with $u<19.$  Fainter than this, the photometric
errors in $u$ will cause increased mixing of the populations.

\section{Detection of a new moving group in the Galactic halo}

We searched our BHB catalog for stars that clustered in velocity, apparent
magnitude, and Galactic coordinates.  Our search was less systematic than that used
by \citet{2006MNRAS.371L..11C}, but allows us to find moving groups that are
more extended, such as tidal streams.  Because they first searched for pairs of 
co-moving stars that were within $2$ kpc of each other, they reduced their sensitivity
to tidal debris streams that could be spread over tens of kpc.

We first divided the catalog into sky areas in $(l,sin(b))$ that were $20^\circ$ by $0.2$
in the respective coordinates,
and made plots of $V_{\mbox{\it gsr}}$ vs. $g_0$ magnitude.  We used $sin(b)$ so that each
partition of the data would cover an equal area of the sky.  Here, $V_{\mbox{\it gsr}}$ refers to the
line-of-sight velocity transformed to the Galactic standard of rest using the Solar 
motion of $(v_X,v_Y,v_Z)=(10,225,7)$ km s$^{-1}$ \citep{dehetal98}.  By examining the
line-of-sight velocity as a function of magnitude, we found a number of
co-moving groups of stars that were associated with known globular clusters, dwarf galaxies, and tidal
streams.  Halo substructures that had been previously identified were not studied
further.  We concentrated only on the most significant of the remaining BHB
association; others were left for future studies.

Figure 1 shows the $V_{\mbox{\it gsr}}$ vs. $g_0$ plot for the most significant co-moving group
of BHB stars, along with the distribution of these stars in $(l,b)$.  Although 
there is a second cluster of stars at $g_0 \sim 16.5$ in the top left plot, we show in
the lower left plot that these stars are dispersed in position on the sky so we did not
explore that association further.

The moving group we describe here has BHB stars at $g_0=18.9$, $\sigma_{g_0}=0.1$; 
and $V_{\mbox{\it gsr}}=-10$ km s$^{-1}$, $\sigma_{\mbox{\it gsr}}=10$ km s$^{-1}$.  The dispersion is
consistent with the instrumental error in the velocity of 19$^{\rm th}$ magnitude
BHB stars \citep{sirkoa}.  These stars have an average heliocentric radial 
velocity of $<\langle V_r\rangle>=-157\pm4$ km s$^{-1}$,
which rules out any association of these stars with the disc population.
In the direction $(l,b)=(65^\circ,48^\circ)$, the line-of-sight, heliocentric
velocities expected for the thin disc, thick disc, and spheroid components
are $-2$, $-35$, and $-144$ km s$^{-1}$, respectively.  The heliocentric radial
velocity of our moving group is not far from that of the average spheroid star
in that direction ($V_{\mbox{\it gsr}}=0$ for the spheroid and $V_{\mbox{\it gsr}}=-10$ km s$^{-1}$
for the moving group), but the dispersion rules out an association with the
general spheroid population.
The lower right panel of Figure 1 shows that the association of horizontal branch 
stars could not be detected spatially; velocity information was required to identify
the moving group.

In the same area, at $(l,b)\sim(59^\circ,46^\circ)$, another cluster identified from SDSS BHB 
stars was found by \citet{2006MNRAS.371L..11C}.  Although this cluster lies at 
roughly the same (l,b) angular position as the cluster identified in this paper, the estimated 
distance to the Clewley et al. clump is 7.8 kpc, which is much closer to the Sun 
than our moving group, with a horizontal branch brighter than $g_0=16$.  

The parameters of the seven BHB stars in the moving group are presented in Table 1.
The new moving group found in this study is located at $(l,b)\sim(65^\circ,47.5^\circ)$.  The 
stars have apparent magnitude in the $g$ filter of $g_0\sim18.9$ and 
velocities located within $-35 < V_{\mbox{\it gsr}} < 15$ km s$^{-1}$, with an average velocity of 
$<V_{\mbox{\it gsr}}>=-10$ km s$^{-1}$.  A $g_0$ apparent magnitude of $18.9$ 
for BHB stars, with an estimated absolute magnitude of $M_{g_0}=0.45$ \citep{2009ApJ...700L..61N} 
corresponds to a distance of $\sim 50$ kpc from the Sun.  The magnitude dispersion of the 
seven stars is $\sigma=0.10$.  Note that the dispersion in apparent magnitude for these 
stars is $\sigma_{g_0}=0.1$, and drops to $\sigma_{g_0}=0.03$ if the two outliers with 
larger magnitudes (one of which has a high estimated surface gravity) are excluded.  The 
magnitudes are consistent with a structure that has zero depth, but the depth could be 
as large as $\sigma=2$ kpc.

\begin{table*}
 \centering
 \begin{minipage}{140mm}
 \caption{Properties of BHBs in the Hercules Moving Group.}
 \begin{tabular}{@{}lrrrrrrrrrrr@{}}
 \hline
ID & RA & DEC & l & b & $g_0$ & $(u-g)_0$ & $(g-r)_0$ & $V_{\mbox{\it gsr}}$ & $\log(g)$ & $[Fe/H]$ & S/N\\
plate-mjd-fiber & hh:mm:ss.s & dd:mm:ss.s & $^\circ$ & $^\circ$ & mag & mag & mag & km s$^{-1}$ & dex & dex &\\
 \hline
1336-52759-290 & 16:08:33.7 & +38:49:19.7 & 61.8 & 47.6 & 19.00 & 1.22 & -0.21 & 2.0 & 2.75 &  -2.99 & 7.7\\
1335-52824-100$^1$ & 16:06:59.7 & +40:20:39.5 & 64.1 & 47.8 & 18.89 & 1.17 & -0.16 & -4.4 & 2.34 & -1.97 & 8.5\\
1335-52824-167$^1$ & 16:04:32.4 & +40:46:03.1 & 64.8 & 48.3 & 18.85 & 1.27 & -0.17 & -10.0 & 2.91 & -2.86 & 9.8\\
1335-52824-362$^1$ & 16:00:37.6 & +41:55:10.5 & 66.6 & 48.9 & 18.82 & 1.17 & -0.20 & -8.3 & 2.83 & -2.63 & 9.2\\
1334-52764-540 & 15:58:24.7 & +43:04:07.4 & 68.4 & 49.2 & 19.10 & 1.16 & -0.14 & -9.7 & 3.39 & -2.10 & 7.7\\
1335-52824-008$^1$ & 16:09:25.9 & +40:05:30.9 & 63.7 & 47.4 & 18.86 & 1.24 & -0.20 & -19.1 & 2.92 &  -2.27 & 8.1\\
0814-52443-046 & 16:17:17.7 & +44:19:04.7 & 69.7 & 45.7 & 18.89 & 1.16 & -0.14 & -18.7 & 2.62 & -2.42 & 9.1\\
\hline
1056-52764-637 & 16:18:21.5 & +36:54:11.0 & 59.1 & 45.7 & 18.90 & 1.32 & -0.12 &  -5.3 & 2.21 & -2.99 & 9.9\\
%1054-52516-518 & 15:56:43.8 & +40:06:48.5 & 64.0 & 49.8 & 20.32 & 2.82 & 0.57 & -21.0 & $-$ & $-$\\
%1170-52756-070 & 16:12:43.1 & +41:52:24.8 & 66.3 & 46.7 & 20.22 & 1.72 & 0.54 & 5.8 & $-$ & $-$\\
%1337-52767-314 & 16:13:48.7 & +37:42:40.1 & 60.2 & 46.5 & 19.34 & 1.17 & 0.49 & -14.2 & $-$ & $-$\\
\hline
\end{tabular}
$^1$ Highly likely member of moving group.
\end{minipage}
\end{table*}

\section{Surface Gravity Estimates of Faint BHB Stars in SDSS}

The SDSS DR7 SSPP pipeline \citep{SSPP1,SSPP2} generates ten different measures of the surface
gravity for each stellar spectrum, and additionally computes an ``adopted" $\log g$, which is
determined from an analysis of the ten different methods.  For stars with S/N less than 10,
the adopted $\log g$ is set to an error code.

We analyzed histograms of the distribution in $\log g$ of bright and faint A stars in our sample,
looking for a method that showed two peaks: one at low surface gravity (BHB stars),
and one at high surface gravity (BS or main sequence stars).  Several of the methods appeared
to separate bright A stars by surface gravity, and none appeared to separate the stars
with S/N$<10$ very well.  We selected the logg9 parameter, which
is described in general terms in \citet{SSPP1},
in which it is referred to as the ``k24 grid".  The original reference to the method is
\citet{ap06}, where the method is described in great detail.  The method finds the best match
between the SDSS spectrum and a set of model atmospheres from \citet{kurucz93}, using the
$(g-r)$ color index and the normalized spectral fluxes in the wavelength range 
$4400 < \lambda < 5500 $ \AA\ ,   using a resolving power of $R=1000$.  The surface gravities using 
this DR7 logg9 estimator are tabulated in Table 1.
Six of the seven candidate members of the moving group have surface gravities less than
$3.0$, as expected for BHB stars.

Subsequent to our search for spheroid structure using BHB stars, \citet{xue} published a list of 
SDSS DR6 A stars along with measurements of $D_{0.2}$ and $f_m$ measurements.  These are
classical indices used by \citet{1983ApJS...53..791P} to classify BHB stars from the $H_\delta$
line width and depth.  Using this method for determining the surface gravities of A 
stars, only 2 of our candidate A stars appear to be low surface gravity.  There are two
effects that contribute to the small fraction of confirmed BHBs: (1) the performance
of the $D_{0.2}$ indicator is degraded at low signal-to-noise, and (2) we disagree with
the published $D_{0.2}$ values for two of the stars.

In the top panels of Figure 2, we show the performance of the $D_{0.2}$, $f_m$ indicators for the
set of $10,224$ stars from \citet{xue}, and for the subset of 1176 of these stars for which
the S/N of the spectrum is between 7 and 10.  While the bright, high signal-to-noise spectra
that were used by \citet{xue} are separable, the lower signal-to-noise spectra do not have
two clear peaks in the diagram.  We show histograms of $D_{0.2}$ for all of
the stars between the two vertical lines ($0.16<f_m<0.24$) in the top two panels of 
Figure 3.  For all of the stars (most of which are high signal-to-noise), we see two peaks
in the $D_{0.2}$ distribution.  The low S/N sample shows no separation.

We compare the separation using the properties of one line of the spectrum ($H \delta D_{0.2}$
vs. $f_m$) with the SSPP logg9 surface
gravities using photometry and the continuum levels of the spectra in the bottom panels of
Figures 2 and 3.  Figure 2 shows that the logg9 indicator works about as well as the
$D_{0.2}$ indicator for bright stars.  Note that in the lower right panel, for stars with
marginal S/N, the distribution of stars is much wider in the vertical direction, and the width
in the horizontal axis is wider but not dramatically so.  The lower panels of Figure 3
show histograms of logg9 for all of the stars between the two horizontal lines
($15<D_{0.2}<40$).  The brighter star data on the left shows two separate peaks for low and high
surface gravity stars, while the marginal S/N data in the right panel does not show 
that separation.  However, we can see from the correlation between $D_{0.2}$ and logg9 
in the lower right panel of Figure 2 that there is some information on the surface 
gravity of even these lower S/N spectra.

Because we were surprised that the $D_{0.2}$ numbers could be so high, even at low
signal-to-noise, for stars that we expected were BHBs in a moving group, we independently
computed new values for the width of the $H_\delta$ at 80\% of the continuum for the eight
stars in Table 1.  For six of them, we measured widths that were reasonably consistent with
the published values.  However, we measured a width of 23 for 1335-52824-167 instead of
40, and a width of 23 for 1335-52824-008 instead of 33.  With these corrections,
six of the eight stars in Table 1 have $D_{0.2}$ widths less than 30, including all four
of the stars that we will later show are highly likely members of the moving group.

\begin{figure*}
\noindent
\centerline{\includegraphics[width=0.90\textwidth]{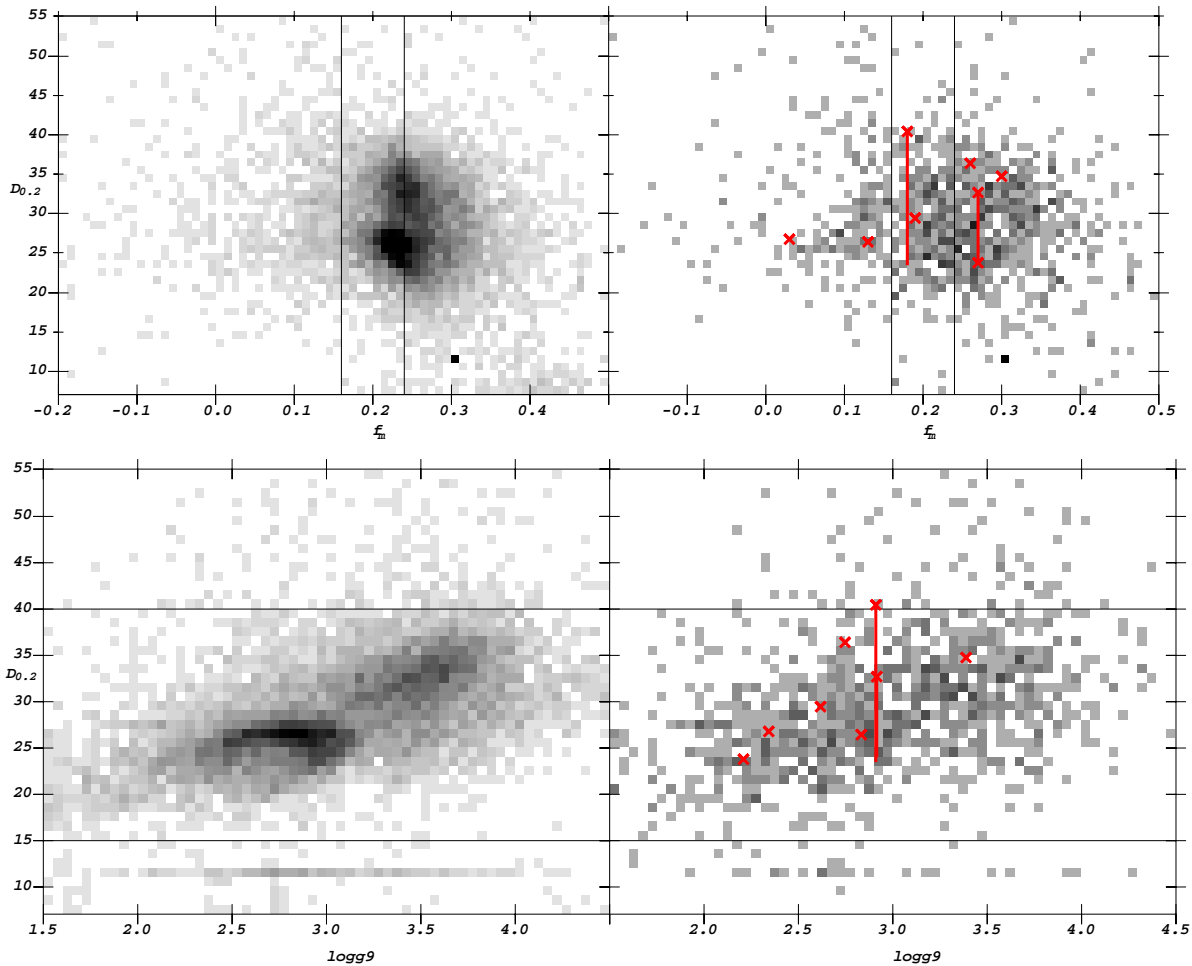}}
%\centerline{\includegraphics[width=0.90\textwidth]{figcombine.ps}}
%\vspace{11pc}
\caption{Luminosity separation of A colored stars into BHB and BS populations.  Upper left panel 
reproduces the sample of 10,224 A colored stars from \citet{xue}, showing that BHBs have
lower $H\delta$ $D_{0.2}$ widths than BSs.  The upper right panel 
shows a low S/N subset of 1,176 stars ($7 < \rm S/N < 10$).  The candidate 
stream members from Table 1, which all have S/N in this range, are indicated with 
(red) crosses.  The single $D_{0.2}$ method appears to be a less effective discriminant at
S/N $< 10$.  However, our independent measurement of $D_{0.2}$ for the moving group
candidates resulted in significantly lower $H_\delta$ widths for two of the eight stars 
(indicated be the vertical line below two of the crosses), so the effectiveness of the
$D_{0.2}$ may depend on how it's measurement is implemented for low signal-to-noise.  The 
lower left panel shows $D_{0.2}$ versus the 9$^{\rm th}$ SSPP method (logg9)for the same 
\citet{xue} sample.  Note that the logg9 surface gravity discriminant achieves similar 
results to the $D_{0.2}$ measure for bright stars.
On the lower right, the lower S/N sample is shown.  A cut at $\rm logg9 < 3.15$ maintains the 
$D_{0.2}$ separation into BHBs and BSs which appears effective
even at S/N $< 10$.  In this paper, we use an even more conservative cut at logg9$<3.0$.  Using this 
conservative selection criterion for BHB stars, 7/8 stream
candidates (red crosses) are consistent with low surface gravity BHB stars.
}
\end{figure*}

\begin{figure*}
\noindent
\centerline{\includegraphics[angle=-90,width=0.95\textwidth]{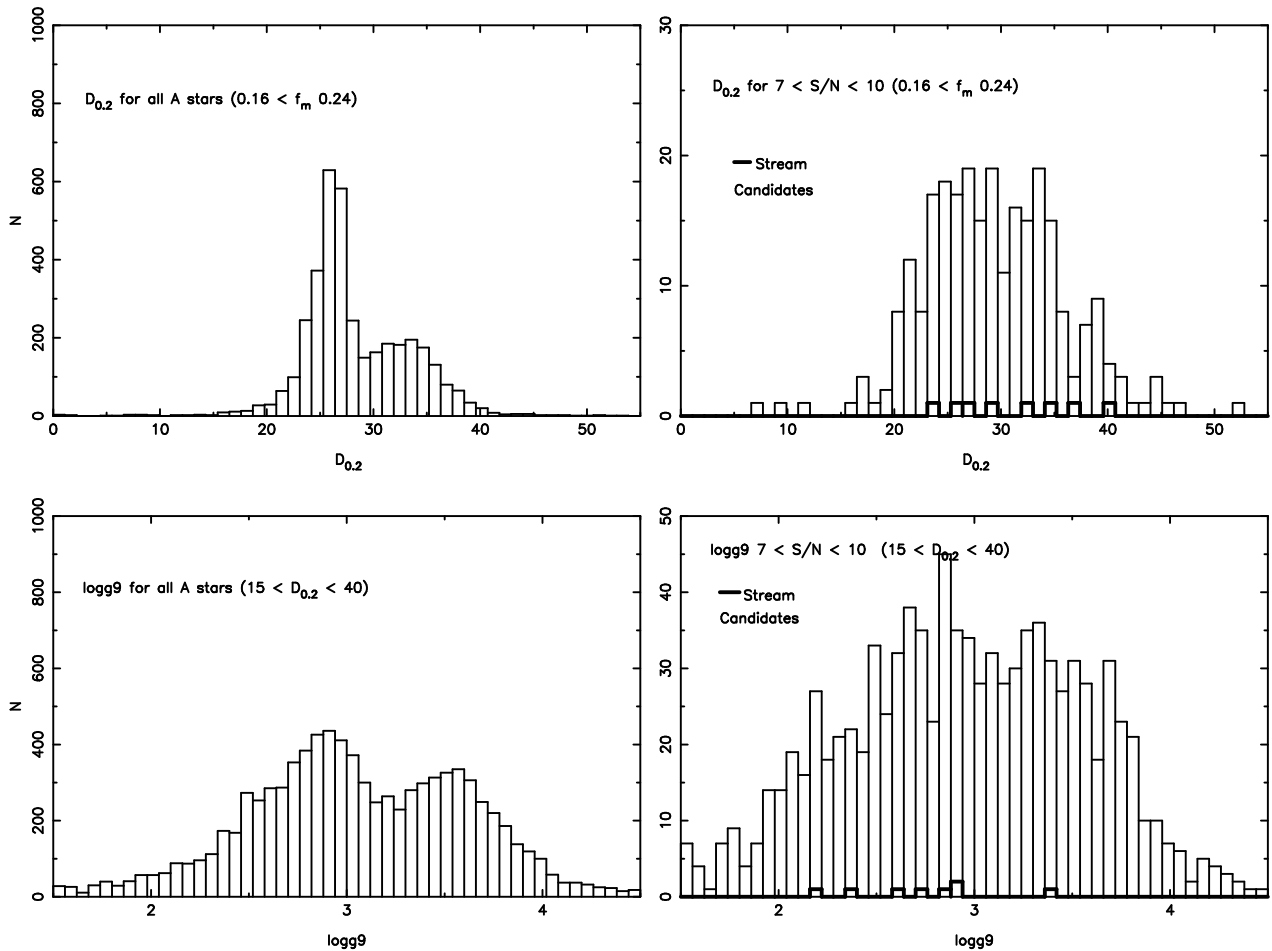}}
%\centerline{\includegraphics[angle=-90,width=0.95\textwidth]{fighist4.ps}}
\caption{Histograms showing gravity separation of data between the parallel lines in each panel of Figure 2.  
Upper left: We show $D_{0.2}$ for the subset of the \citet{xue} sample with an $H\delta$ flux minimum (as 
fraction of continum) of $0.16 < f_m <0.24$.  
Upper right: We show the same histogram for the lower ($7 < \rm S/N < 10$) signal-to-noise subsample.  Stream
candidates from Table 1 are indicated with a heavy curve. Note the two values at 40 \AA\ and 33 \AA\ should be moved lower (23 \AA\ ) based on inspection of individual spectra. Lower Left: We show the distribution of 
logg9 for most of the \citet{xue} sample.  Lower Right: We show the lower S/N subsample.  Note 
the positions of the stream candidates (dark line), which are consistent with the low surface gravity BHB population 
in 7/8 cases.  Note the clear bimodality between BHBs (low $D_{0.2}$ or low logg9) and BSs
(high $D_{0.2}$ or high logg9) for the full samples, and the absence of bimodality for the lower S/N
samples.  
}
\end{figure*}

\section{Statistical Significance of the Moving Group}
Given that we were searching through the data by eye and looking for clusters, it is
important to verify that the clustering of stars is not a chance coincidence.  In this section
we will estimate an upper bound for the probability that this clump could be a
random fluctuation in the number of stars in a a small region of angle in the sky,
magnitude, and line-of-sight velocity.

Most of the sky that we searched for clumps was at high Galactic latitude; at low Galactic
latitude the stellar population is different, the density of stars is higher, and the sky
coverage is less uniform.  So for statistical purposes we limited our sample to $b>35^\circ$.
There are 2893 stars in our sample with $b>35^\circ$.  The seven stars were discovered
in a 19 square degree area of the sky (compared with 8800 square degrees above $b>35^\circ$), 
with the requirement that this sky area be rectangular
in Galactic latitude and longitude ($61.8<l<69.7, 45.7<b<49.2$).  In comparison with the entire
high latitude sample of BHB candidates,
the seven stars in the clump span about 1/46$^{\rm th}$ of the entire longitude range
of the input data, and 1/10$^{\rm th}$ of the latitude range of the data in an equal-area plot.
From Figure 1, we find that the range of velocities in the clump includes $15\%$ of the 
sample stars, and the range of magnitudes in the clump is the same as $10\%$ of those in the 
high latitude sample.  

Appendix A presents a statistical method to determine the significance of a ``clump" of data 
points in a $d$-dimensional search space.  In this case we have four dimensions: Galactic
longitude, Galactic latitude, radial velocity, and apparent magnitude.  The statistical
method measures the probability that random data will produce a clump of seven or more
points within a four-dimensional box of the measured size.  By using percentiles, we account
for the non-uniformity of the data in the radial velocity and apparent magnitude dimensions.
We also adjust the statistics, as explained in Appendix A.4.2, for the periodic boundary 
condition in Galactic longitude.  What is important for the calculation is that each of 
the dimensions be independent (separable, in the terms of Appendix A).  In our case, the 
variables are not quite independent, especially Galactic coordinates.  Because the density 
of A stars with SDSS spectra varies as a function of position in the sky, the relative percentiles 
in $(l,b)$ would be different if a clump of the same areal extent were discovered in 
a different part of the sky.  We estimate that the local density in the region of the sky
that the clump was found is 3.5 times higher than average.  Therefore, following the prescription
of Appendix A4.4, we multiplied the longitude width of the clump by a factor of 3.5.

It was not possible to account for every known peculiarity of the data.  For instance, when we
searched for clumps we
divided the sky up into fixed $(l,b)$ boxes, so there are regions around the edge of each
$(l,b)$ box that were not as thoroughly searched.  However, it is possible that if nothing was
found with these limits we would have adjusted the box sizes and positions, so this oversight
(which would underestimate the significance of a clump) is probably justified.  

Given the percentiles of the 4-D search space listed above, we expect 0.32 stars in a typical 
box with these dimensions.  Using the algorithm in Appendix A, the number of clumps we should
expect to find in this data that have seven or more stars in a 4-D box of these dimensions 
or smaller is 0.96.  Since we found one clump, this is very close to random.  
If we eliminate one
star from the sample that is an outlier in $(l,b)$, so that the area of sky is now
$61.8^\circ<l<68.4^\circ$, $47.4<b<49.2^\circ$, then we calculate that
the E-value for finding six or more stars in a 4-D box
with these (smaller) dimensions is 0.18, so the probability of finding this moving
group is not more than one in six.  Using only these dimensions, the clump of stars is
not a significant detection.

In order to claim a significant detection of the moving group represented by these seven
stars, it is necessary to compare their metallicities with the metallicities of the
background population.  Only 30\% of the stars in the sample have metallicities as low as
those in the clump of seven stars.  If we extend the statistical calculation to five
dimensions, including metallicity, then the probability of finding seven stars in a region
of this size and shape (or smaller), anywhere in the 5-D parameter space is not more than
1 in 244 (E-value of 0.004), and the probability of finding six stars in a smaller 
angular region in the sky (in 5-D) is not more than 1 in 490 (E-value=0.002).  These are 
highly significant detections.

Since metallicity is critical to the detection of this substructure, we examine the SDSS DR7
metallicity measurements of these stars, and BHB stars in general, in the next two sections.  In
section 8, we use the information we have learned in the analysis of this moving group to
re-select similar data from SDSS DR7, and make an even stronger argument for the statistical
significance of our result.

\section{Metallicities of the moving group stars}

All seven of the candidate BHB stars in this moving group had unmeasured metallicities in
SDSS DR6, but metallicities were assigned to these same spectra in the much improved SSPP pipeline
of SDSS DR7 \citep{SSPP1,SSPP2}.  Our metallicities come from the DR7,
which became public in October 2008.  The SDSS DR7 tabulates many measures
of the metallicities of stars, of which the most commonly used is FEHA, the
``adopted" metallicity, which is derived from a comparison of all methods
used to measure metallicity.  In this paper we use instead the metallicity of 
\citet{wbg99}, hereafter WBG, which is specifically designed to measure the metallicities
of BHB stars.  The WBG metallicities of the seven BHB stars
in the moving group are given in Table 1 and are shown in Figure 4.  The mean of the 
distribution is [Fe/H]$=-2.46\pm0.14$, and the width is $\sigma=0.4$.

\begin{figure}
%\plotone{fig5.ps}
\noindent
%\centerline{\includegraphics[angle=-90,width=.45\textwidth]{fig5.ps}}
\centerline{\includegraphics[angle=-90,width=.45\textwidth]{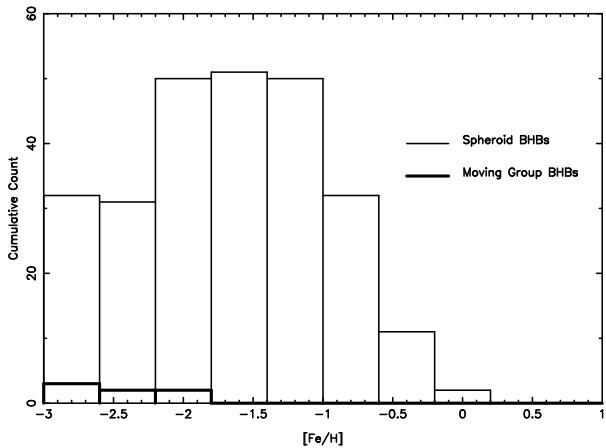}}
%\vspace{11pc}
\caption {Metallicity distribution of the moving group.  We selected all stars with spectra from SDSS DR7.1 which had
the same photometric properties as the moving group BHB stars ($18.65 < g_0 < 19.15$, $-0.3 < g-r_0 < 0.0$,
$0.8 < (u-g)_0 < 1.5$, and split photometrically by likelihood of low surface gravity), with Galactic
latitude $b>45^\circ$.  The 298 stars selected are all likely spheroid BHB stars, of which 260 had measured
metallicities.  The distribution of metallicities 
of the spheroid BHB stars with similar photometric properties (thin lines) is shown along with the metallicities of the
seven BHB stars in the moving group (thick lines).  The moving group has a mean metallicity of $-2.5$, while the
spheroid BHBs have a mean metallicity of $-1.69$.
\label{metals}}
\end{figure}

We now show that the metallicities of the moving group stars are not consistent with
being drawn at random from the stellar halo.
We selected all stars in DR7 with the same
photometric constraints as the stars in the moving group, and with $b>45^\circ$ so that the
BHB stars are likely to come from the same component (the spheroid), and not be confused with
disc populations of BHBs.  To remove BHB stars in the Sagittarius dwarf tidal
stream from the sample, we also eliminated stars with $\delta<5^\circ$.  
There were 298 stars with $18.65 < g_0 < 19.15$, $-0.3 < (g-r)_0 < 0.0$,
$0.8 < (u-g)_0 < 1.5$, and that passed the photometric low surface gravity cut.  Of these, 260 had measured
metallicities that are shown in Figure 4.  The mean of the distribution is
[Fe/H]$=-1.69\pm0.04$, and the width is $\sigma=0.71$.  Note that the error quoted
here is a statistical error.  Because very few of the SEGUE calibration stars
are as blue as A stars, the systematic errors at this end are about $\pm 0.3$.
The comparison stars are about 45 kpc from the Sun, and above $b=45^\circ$, so
they come from the distant halo.  The blue straggler contaminants are 10 to 28
kpc away, which is also fairly far from the disc at high Galactic latitudes.

A comparison of the stars in the moving group with the similarly selected stars
in the spheroid with a t-test gives a probability of 1 in $10^6$ that the two groups
of stars were selected from the same stellar population.  The Mann-Whitney
test (also known as the Wilcoxon rank sum test) gives a p-value of 0.00023, or a 
2 in 10,000 probability that the two samples are drawn from the same population.
This leads us to conclude that the moving group is real.

As a further test, we selected the other 8 stars in the lower panel of Figure 1 that have the
same $V_{\mbox{\it gsr}}$ and apparent magnitude as the blue horizontal branch stars in the
moving group, and find that they have a mean metallicity of $-1.87$ with a sigma of 0.86. 
The t-test (p=0.948) shows that this distribution is not
distinguishable from the background population.  (Performing the Mann-Whitney
test for this sample is problematic since several of the stars are not
part of our selected background distribution.)

\section{Systematics of DR7 Metallicities of halo BHB stars}

\begin{figure}
\noindent
\centerline{\includegraphics[angle=-90,width=0.45\textwidth]{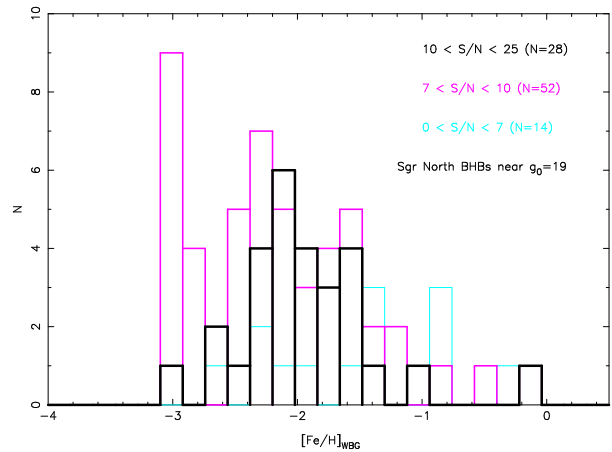}}
%\centerline{\includegraphics[angle=-90,width=0.45\textwidth]{figfeh.ps}}
\caption{Measured metallicities of BHB star in the Sgr dwarf leading tidal tail, as a function
of S/N.  The metallicities of stars with S/N$>10$ are considered fairly reliable.  The stars
with $7<$S/N$<10$ have a similar distribution.  Stars with $0<$S/N$<7$ have a significantly
higher average metallicity.  Therefore, we conclude that the metallicity measurements of 
BHB stars with S/N$<7$ are not reliable.  Note that the excess of stars at [Fe/H]$_{\rm WBG}=-3.0$
is due to edge effectes in the WBG estimation procedure, and does not represent a cluster
of stars with identical metallicity.
}
\end{figure}

In this section we explore the accuracies of the DR7 metallicity measurements of BHB stars
as a function of S/N and in comparison with published metallicities for globular clusters.
We also explore the metallicity of BHB stars in the halo as a function of apparent magnitude.

Since all of our moving group stars have low S/N, we need to test the accuracy of the metallicity
measurements.  We do this by selecting BHB stars in the leading tidal tail of the Sgr dwarf galaxy
that are about the same distance from the Sun as the moving group.  We selected all of the spectra 
from SDSS DR7 that had elodierverr$>0$ (if the radial velocity error when matching to the stellar
templates is less than zero, that indicates that the object is not a star), $0.8<(u-g)_0<1.5$, 
$-0.3<(g-r)_0<0.0$, $320^\circ<l<350^\circ$,
$30^\circ<b<70^\circ$, $-100<V_{gsr}<100$ km s$^{-1}$, logg9$<3.0$, and $18.5<g_0<19.6$.
The 94 stars that are likely BHBs in the Sgr dwarf tidal stream were divided into three files based
on S/N.  There were 14 stars with $0.0<$S/N$<7$, 49 stars with $7<$S/N$<10$ (excluding three with
unmeasured metallicities), and 28 with $10<$S/N$<25$.  The metallicity distributions of these three
S/N groups is shown in Figure 5.

The highest S/N group ($10<$S/N$<25$) has a mean [Fe/H]$_{\rm WBG}=-1.94$, which is similar to the
group with marginal S/N ($7<$S/N$<10$) which has a mean [Fe/H$_{\rm WBG}=-2.13$.  A Mann-Whittney
test comparing these two samples gives a result of $U=818$, $z=-1.39$, and a p-value of
0.165.  These distributions are not significantly different from each other.  This gives us some
confidence that the metallicities of the moving group stars, which have S/N measurements in 
this marginal range, are usefully measured.

In contrast, the lowest S/N group ($0<$S/N$<7$) has a mean of [Fe/H]$_{\rm WBG}=-1.42$.  A
Mann-Whittney test comparing this sample with the highest S/N sample produces $U=275.5$,
$z=-2.11$, and a p-value of 0.0349.  This is a statistically significant difference,
and metallicities of BHB stars measured by this technique are probably not accurate.

To understand the likely systematics in our metallicity determinations, we 
selected BHB stars from six globular clusters that had spectroscopic measurements 
of BHB stars in SDSS DR7.  We selected stars that have the colors of BHB stars,
are within half a degree of the centre of the globular cluster, and have
radial velocities within 20 km s$^{-1}$ of the published value for the cluster.
M2, M3, and M13 have metallicities very close to $-1.6$; M53 has metallicity of 
$-1.99$; and the two clusters M92 and NGC 5053 have metallicities very close to
$-2.3$ \citep{harris}.  Because there were very few stars with spectroscopy in
each globular cluster, we combined the data from clusters with similar
metallicities before plotting the histograms in Figure 6.  All of these
globular clusters have horizontal branches near $g_0=16$ (the magnitude
range for which we have the most complete sample of BHB stars), and thus
the spectra have high $S/N$.

\begin{figure}
%\plotone{figmeta.ps}
\noindent
\centerline{\includegraphics[width=.45\textwidth]{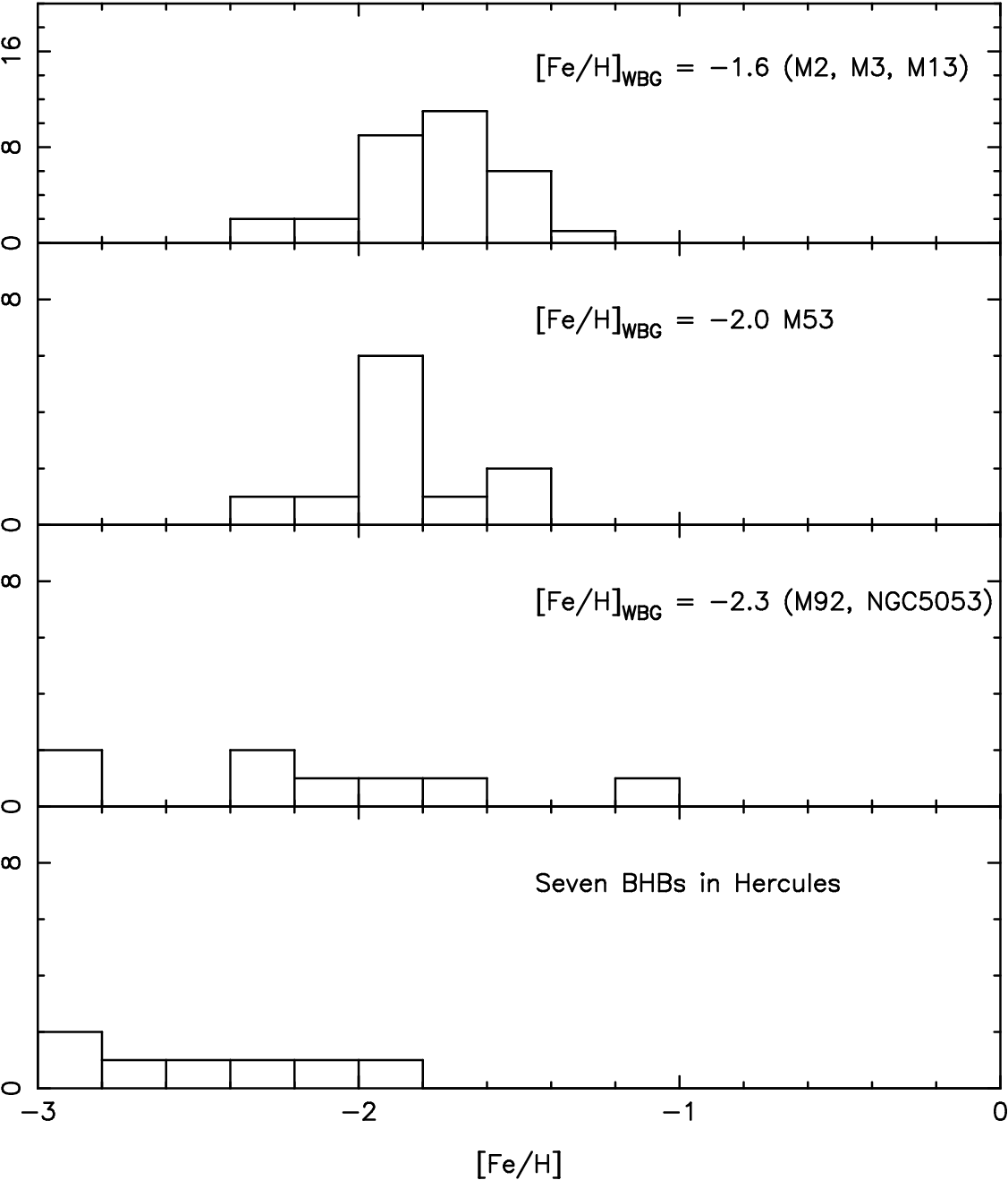}}
%\centerline{\includegraphics[width=.45\textwidth]{figmeta.ps}}
%\vspace{11pc}
\caption{Metallicity measures of known clusters.
We show the metallicity distribution of SDSS DR7 BHB stars, as 
measured by the \citet{wbg99} technique, for BHB stars selected from
clusters of known metallicity.
Uppermost panel: Stars in M2 ([Fe/H] = $-1.62$), M3 ([Fe/H] = $-1.57$), and 
M13 ([Fe/H] = $-1.54$).
Second panel: Stars in M53 ([Fe/H] = $-1.99$).
Third Panel: Stars in M92 ([Fe/H] = $-2.28$) and NGC 5053 ([Fe/H] = $-2.29$).
Lowermost Panel: The seven Hercules stream BHBs.
Note that the measured metallicities are compressed toward [Fe/H]=$-1.9$
(higher metallicity stars are measured systematically low, and lower
metallicity stars are measured systematically high).  Also note that
the Hercules stream stars appear to be lower metallicity than M92 and
NGC 5053.
}
\end{figure}

Figure 6 shows that the SDSS DR7 WBG metallicities for BHB stars tend to
push the metallicities toward [Fe/H]=$-1.9$.  The measured metallicities
are correlated with the published metallicities of the globular clusters;
however, stars that have higher
metallicities are measured with metallicities that are too low, and stars
with lower metallicities are measured systematically too high.

We show for comparison the measured metallicities of the seven stream stars.
The measured metallicity is lower than that of M92 and NGC 5053.

Because the validity of our stream detection depends on our ability to separate
the stream from the stellar halo in metallicity, we need to show that the
metallicities in our background population are accurate.  We selected all
of the stars with spectra in DR7 that have photometry consistent with being
a BHB star, as explained in \S 2, $b>45^\circ$, and $\delta>5^\circ$.  We 
further restricted the sample by
insisting that the WBG estimate of $\log g$, as measured in DR7, was less
than 3.75.  

In Figure 7, we show the DR7 WBG metallicity as a function of apparent
magnitude.  We have spectra for stars with $14.5< g_0 <19.15$, which span
distances of 6 to 50 kpc from the Sun, all at high Galactic latitude.
The metallicity distributions for all of the stars brighter than $g_0<18.5$
are similar, with a mean near [Fe/H]=$-1.9$ and a sigma of 0.45.  In the
faintest set of stars, which have apparent magnitudes similar to that of the
newly detected stream, the distribution appears considerably broader, but
with a similar mean.  We attribute the increased width to the lower 
S/N in many of these spectra.  The mean is somewhat lower in the last
panel in Figure 7 than in Figure 4, probably due to a cleaner sample of BHB stars, with
less contamination from BS, from the additional restriction in 
$\log g$.

\begin{figure}
%\plotone{figfeh10.ps}
\noindent
\centerline{\includegraphics[width=.45\textwidth]{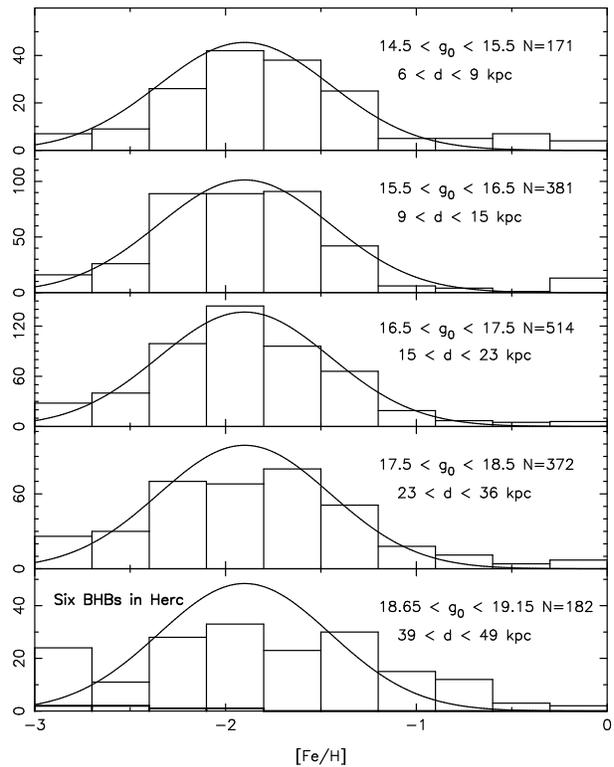}}
%\centerline{\includegraphics[width=.45\textwidth]{figfeh10.ps}}
%\vspace{11pc}
\caption{Metallicity of the outer halo vs. distance.
We show the metallicity distribution of SDSS DR7 BHB stars in several
apparent magnitude ranges.  
The last panel shows the six Hercules stream BHBs (thick lines in lower
left of diagram) that survive the
additional cut in $\log g$.  They are inconsistent with being drawn from
the same distribution as the other stars in this panel.
All of the stars have $ugr$ colors consistent
with BHB stars, $b>45^\circ$, $\delta>5^\circ$, and $\log g <3.75$.  The
[Fe/H] and $\log g$ measurements are from the WBG \citep{wbg99} techniques,
as implemented in the SSPP of DR7.  A Gaussian with mean $-1.9$ and sigma
0.45, normalized to the number of stars in each panel, is shown for
reference.  The mean metallicity does not change as a function of distance
from the Sun.  The distribution is wider in the lowest panel because the
signal-to-noise is lower here.  If one selects only the highest S/N
spectra in this bin, the width is similar to the four other panels.
}
\end{figure}

We show for comparison the six stream stars that have $\log g < 3.75$ in the
last panel of Figure 7.  The metallicity distribution of these stars is
inconsistent with the distribution of the other stars in this figure at
the 98\% confidence level, as determined by the Mann-Whitney test.

The value that we find for the metallicity of the outer halo, as measured
from SDSS BHB stars is on the low side of the distribution of previous
measurements, but higher than the \citet{carollo} metallicity measurement
of [Fe/H]=$-2.2$.  (We note, however that Carollo et al. 2007
distribution of BHB stars in their paper, and the BHB star distributions
peak closer to [Fe/H]=$-2.0$.)  

\section{The Moving Group in DR7}

\begin{figure}
\noindent
\centerline{\includegraphics[angle=-90,width=0.45\textwidth]{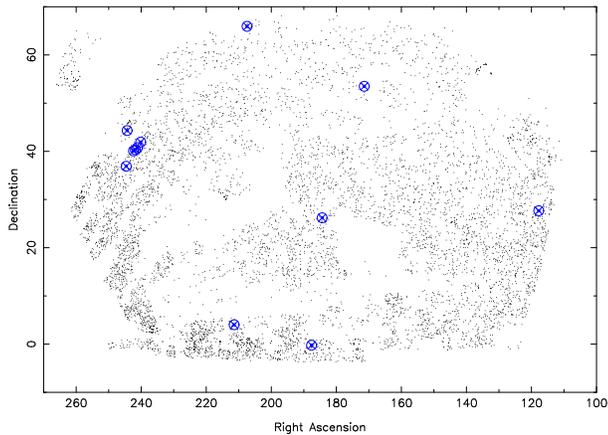}}
%\centerline{\includegraphics[angle=-90,width=0.45\textwidth]{figradec.ps}}
\caption{Projection  of the moving group in equatorial coordinates, as detected in SDSS DR7.
The small black dots show the positions of A stars with spectra that are not from SEGUE plates
as selected from DR7.  Only twelve of these stars (large cyan symbols) have surface 
gravities, metallicities,
apparent magnitudes, and line-of-sight velocities that are consistent with the moving
group, as detected in Figure 9.  Of these twelve, four are very tightly clustered and a fifth
is only a few degrees away and in line with the other four.
}
\end{figure}

\begin{figure}
\noindent
\centerline{\includegraphics[angle=-90, width=0.45\textwidth]{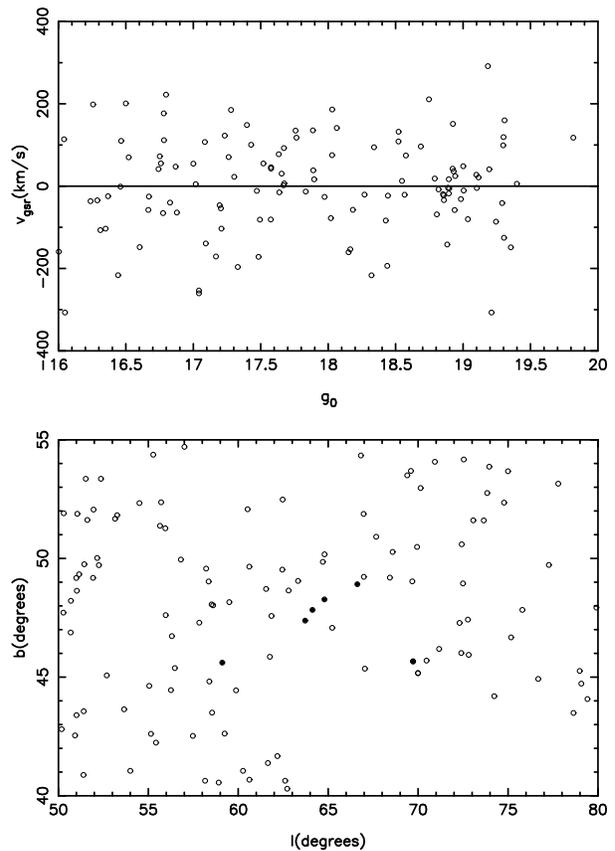}}
%\centerline{\includegraphics[angle=-90, width=0.45\textwidth]{fig2panel.ps}}
\caption{Detection of the moving group in SDSS DR7.  This figure is similar
to the upper right and lower left panels of Figure 1, with the additional restriction that the all the stars in these panels have $\rm [Fe/H]_{\rm WBG} < -1.95$.  The significance of the group over background is enhanced with this
metallicity cut.
}
\end{figure}

The moving group was originally discovered by searching through data from SDSS DR6.
In this section we
show the moving group as selected from the now public SDSS DR7 databases, and
using our knowledge of metallicity and surface gravity determination at low S/N,
as learned in the production of this paper.

We selected from SDSS DR7 all of the spectra with elodierverr $>0$ (which selects
stellar spectra), magnitude $15<g_0<23$, the colors of an A star
[$0.8<(u-g)_0<1.5, -0.3<(g-r)_0<0.0$], and a flag that indicates the spectrum 
is not from a SEGUE plate.  We did not select stars from SEGUE plates
since most of them have not been searched for moving groups of A stars, and because
their non-uniform sky coverage, different section criterion, and different exposure
lengths make it difficult to use them in a statistical calculation. 
We show the distribution of SDSS DR7 A stars in equatorial coordinates in Figure 8.

We then selected the low surface gravity (logg9 $ <3.0$) stars with $50^\circ<l<80^\circ$ 
and $40^\circ<b<55^\circ$,
to match the region of the sky presented in Figure 1, and metallicity [Fe/H]$_{\rm WBG}<-1.95$,
since we know we are looking for a low metallicity moving group.  Figure 9 shows plots like
those used in the moving group discovery, but using data from this low metallicity DR7
dataset.  From the upper panel, we identify a clump of stars with $-35<V_{gsr}<0.0$ km s$^{-1}$
and $18.8<g_0<18.9$.  The lower panel shows the positions in Galactic coordinates of the
same objects from the top panel, with filled circles showing the six objects that are clumped
in line-of-sight velocity and apparent magnitude.  Four of these objects are very tightly
aligned in angular position on the sky.

To get a sense for how common a moving group with these characteristics is, we selected from
our original DR7 dataset all of the stars with surface gravity, metallicity, line-of-sight
velocity, and apparent magnitude that are similar to the stars in the clump.  In the entire
northern sky portion of the SDSS, there are twelve stars with these characteristics.  Four of
the stars are in a tight line on the sky, and a fifth is only a few degrees away from the clump
along the same line.  Because this fifth star is a candidate member of the moving group, and
not included in the original list of seven, we have added it to the bottom of the list of
candidate members in Table 1.  The four highly probable members of the moving group,
which were selected from both the DR6 and DR7 datasets, are marked in Table 1 with a footnote.

The four highly likely moving group members are a subset of the original seven, and have 
velocity and metallicity characteristics that are nearly identical to our previous results.  The 
average $V_{gsr}$ is $-10 \pm 3$ km s$^{-1}$, with a sigma of $6$ km s$^{-1}$.  The 
average metallicity is [Fe/H]$_{\rm WBG}=-2.4 \pm 0.2$ with a sigma of 0.39.

Following a similar procedure to that used in Section 5, we now calculate the probability
that this group of four stars is a random coincidence.  There are 6118 stars with $b>35^\circ$
in the A star sample depicted in Figure 8.  Of these stars, 3335 have $-9<{\rm logg9}<3.0$
(54.5\%),  1171 have $-9<$ [Fe/H]$_{\rm WBG}<-1.95$ (28.9\%), 224 have $18.8<g_0<18.9$
(3.7\%), and 487 have $-23<V_{gsr}<0$ km s$^{-1}$ (8.1\%).  The four stars in the clump have
sky positions within $63.72^\circ<l<66.62^\circ$ (0.8\% of longitude range) and 
$47.37^\circ<b<48.92^\circ$ (4.2\% of area-corrected latitude range).  Because the density 
of the stars in Figure 8 is not uniform, we need to use a multiplier on the fraction of stars
in the longitude range, since that is fractionally the smallest dimension.  In the
3.0 square degree area that contains the four stars, there are 8 stars in the original sample
of 6118.  So, in the region of the clump there are 2.7 stars per square degree, while there
are $6118/8796=0.69$ stars per square degree in the whole sky region, a factor of 3.9.  Using
the procedure outlined in Appendix A, we calculate that we would expect 0.0038 stars in a 6-D
region that instead has four stars.  The expected number of clumps of four stars is
0.0238.  The p-value is therefore less than 0.024 ($p<E=0.238$).  Since $p<0.05$, this is a 
significant detection.

Note that this calculation is extremely conservative.  First, the calculation assumed that we
searched all 6118 stars for clumps, when in fact we only searched the ones that had low
surface gravity, as determined by the photometric indices.  Second, the four stars
are not randomly distributed within the $(l,b)$ box, they are tightly constrained to a line,
as one would expect for a tidally disrupted globular cluster; if our statistical analysis
allowed a non-axis aligned region, the computed statistical significance would have been
stronger.  And third, when we calculated
the density of stars in the region containing the moving group we included the moving group
stars (half of the sample of eight) in the density calculation.  This artificially inflates our 
Galactic latitude multiplier.  To estimate the effect of just this third factor, we counted
the number of A stars in a slightly larger region of the sky with $60^\circ<l<66.63^\circ$
and $47.37^\circ<b<48.92^\circ$ and found 16 stars, of which 4 are members of the moving group.
This gives us a local density of $12/6.87=1.75$ stars per square degree, which is a factor
of 2.5 times the average star density.  Using this multiplier, we expect to find 0.0025
objects where we actually find four, and the expected fraction of the time a clump of four 
stars in this size region of 6-D space should be found in this size data set is not more
than 0.0063.

\section{Estimated luminosity of the progenitor}

We estimate the properties of a progenitor of this moving group, making the assumption 
that it was a globular cluster, and that we have detected the 
majority of the BHB stars from what was the core of the star cluster.  We expect the
progenitor was a globular cluster because the stars are very well aligned in the sky, have
a velocity dispersion consistent with instrumental errors, and there are very few stars
identified.  If the progenitor
was a dwarf galaxy, so that it also has younger or more metal rich components, or if
we have found only a knot of increased stellar density along a longer tidal debris stream,
then the inferred size of the progenitor is larger.

We detected 4-8 BHB stars in the moving group.  In the color-magnitude range in the vicinity
of the moving group, we have spectra of 50 of the 73 candidate BHBs (68\%).  Therefore,
we expect that there are $\sim 10$ BHBs in the moving group.  \citet{1997A&AS..121..499B} found
five BHB stars in the globular cluster Pal 13 ([Fe/H]=$-1.74$, $M_V=-3.74$, Harris 1996).  
\citet{ynetal00} find at least five BHB stars in the globular cluster Pal 5 
([Fe/H]=$-1.41$, $M_V=-5.17$, Harris 1996).  The moving group is consistent with a
progenitor that is like one of the smaller globular clusters of the Milky Way, with
a total integrated luminosity of about $M_V\sim -4$.

To understand why this moving group could not be identified from photometry alone, we estimated
the number of red giant and subgiant branch stars in the moving group by comparison with
NGC 2419.  Note that at 50 kpc, turnoff stars are at $g_0 \sim 22.7$, which is close to the
limiting magnitude of the SDSS, so deeper photometry is required to clearly detect
significant density enhancements expected from turnoff and main sequence stars
fainter than these limits.  We selected SDSS DR6 stars within about nine arcminutes 
of the centre of NGC 2419 (this does not include stars near the very centre of the GC,
since individual stars are not resolved there).  BHB stars were selected with $20.2<g_0<20.6$
and $-0.4<(g-r)_0<0.0$.  Giant stars were selected within a parallelogram with vertices
$[(g-r)_0,g_0]=[0.5,20], [0.9, 18], [0.7, 20], [1.0, 18]$.  Subgiant stars were selected
within a triangular area with vertices $[(g-r)_0, g_0]= [0.5, 21], [0.1, 23], [0.6, 23]$.
We found 101 BHBs, 98 red giants, and 302 subgiants over background in the region of sky
that was searched.  Therefore, we expect about the same number of red giant stars as BHBs
in our moving group, and about three times as many subgiants.  We shifted the magnitudes
of the color-magnitude boxes by 1.41 magnitudes (the difference in distance modulus between
our moving group and NGC 2419) and counted the number of stars in
$60^\circ<l<70^\circ$ and $45^\circ<b<50^\circ$.  The moving group is expected to have
10 of 73 candidate BHB stars (1 sigma fluctuation), 10 of 2742 candidate giant stars (0.2 sigma), 
and 30 of 13,182 candidate subgiant stars (0.3 sigma).  The background star counts are much 
too high for us to find this moving group in photometry alone.

%It should be noted, however, that the case that the moving group is a narrow tidal
%stream, like that of the Sagittarius dwarf spheroidal galaxy or the Pal 5 globular cluster,
%is weak.  The stars are not localized in a small cross section like a globular cluster
%tidal stream and the density of stars is low compared to previously detected dwarf
%galaxies.  Although it is likely that there is an association between at least some of
%these seven stars, the evidence that the association is limited to the small region studied
%is less solid.  It is possible that the stars originated in a larger cluster of stars
%that was dispersed early in the history of the Milky Way, and has been spread across the
%sky.

\section{Conclusions}

We have detected a moving group of at least four BHB stars in the corner (``toe") of the 
Hercules constellation spilling into Corona Borealis.  These stars are coincident in angular position 
[$(l,b)=(65^\circ,48^\circ$)], apparent
magnitude ($g_0=18.9$), line-of-sight velocity ($V_{\mbox{\it gsr}}=-10$ km s$^{-1}$, 
$\sigma_{V_{\mbox{\it gsr}}}<10$ km s$^{-1}$, $\langle V_r\rangle=-157$ km s$^{-1}$), and 
metallicity ([Fe/H]=$-2.4$).  We expect that the progenitor of this moving group was a low
metallicity globular cluster, with a luminosity like that of one of the smaller globular
clusters in the Milky Way halo.

We show that useful surface gravities and metallicities
are measured for BHB stars with S/N$>7$ in SDSS DR7.  The mean 
metallicity of BHB stars in the outer halo is similar to M53, which has a published 
metallicity of [Fe/H]=$-2.0$.  The metallicity does not appear to change with distance from
the Sun ($6<R<55$ kpc).  Our measurement of the spheroid metallicity is slightly higher 
than claimed by \citet{carollo} and somewhat lower than earlier studies 
of outer halo stars.  
%We have identified several additional stars with the same
%velocities with photometric properties that would be expected from G subgiant stars with
%[Fe/H]=-2.0.  

The Hercules moving group is one of many tidally disrupted stellar associations 
expected to comprise the spheroid of the
Milky Way and could not have been identified from photometry alone; more 
complete spectroscopic surveys are required to identify the component
spheroid moving groups, and determine the merger history of our galaxy.
We present a statistical technique that allows us to estimate the significance of clumps
discovered in multidimensional data.

\section*{Acknowledgments}

This project was funded by the National Science Foundation under grant number AST 06-07618.
T.C.B and Y.S.L. acknowledge partial support from grans PHY 02-16783 and PHY
08-22648, Physics Frontier Centers/JINA: Join Institute for Nuclear Astrophysics,
awarded by the National Science Foundation.  P.R.F. acknowledges support through the Marie
Curie Research Training Network ELSA (European Leadership in Space Astroometry) under
contract MRTN-CT-2006-033481.
Many thanks to Ron Wilhelm, who answered our questions about potential RR Lyrae stars and
stellar metallicities.

Funding for the SDSS and SDSS-II has been provided by the 
Alfred P. Sloan Foundation, the Participating Institutions, the 
National Science Foundation, the U.S. Department of Energy, the 
National Aeronautics and Space Administration, the Japanese 
Monbukagakusho, the Max Planck Society, and the Higher Education Funding 
Council for England. The SDSS Web Site is http://www.sdss.org/.

The SDSS is managed by the Astrophysical Research Consortium for 
the Participating Institutions. The Participating Institutions are the 
American Museum of Natural History, Astrophysical Institute Potsdam, 
University of Basel, University of Cambridge, Case Western Reserve 
University, University of Chicago, Drexel University, Fermilab, the 
Institute for Advanced Study, the Japan Participation Group, Johns 
Hopkins University, the Joint Institute for Nuclear Astrophysics, the 
Kavli Institute for Particle Astrophysics and Cosmology, the Korean 
Scientist Group, the Chinese Academy of Sciences (LAMOST), Los Alamos 
National Laboratory, the Max-Planck-Institute for Astronomy (MPIA), the 
Max-Planck-Institute for Astrophysics (MPA), New Mexico State 
University, Ohio State University, University of Pittsburgh, University 
of Portsmouth, Princeton University, the United States Naval 
Observatory, and the University of Washington.

%\clearpage

\appendix

\section{The Statistics of Clumpy Data}

\subsection{Introduction}

For this paper, we needed to estimate the probability of finding seven
BHB stars in a restricted portion of $l,b,V_{\mbox{\it gsr}},g_0$ parameter space.
In this appendix, we present a solution to the more general problem
of determining the probability of finding $K$ of $N$ data points in
a small portion of a $d$-dimensional space.

Suppose one analyses of  a set of $N$ data points in a large $d$-dimensional hypercube
$[A_1, B_1] \times \ldots \times [A_d, B_d]$\@, and finds a small,
axis-aligned, $d$-dimensional hypercube $[a_1, b_1] \times \ldots
\times [a_d, b_d]$ that has $K \ge 2$ data points in it.  The natural
question is whether this ``clump'' of $K$ data points is statistically
significant, or whether even random data would yield such a clump.
Here we precisely estimate an $E$-value for the expected number of
clumps, thus providing an upper bound to the probability that random
data would somewhere have $K$ or more data points within an
axis-aligned, $d$-dimensional hypercube with dimensions $(w_1, \ldots,
w_d) = (b_1 - a_1, \ldots, b_d - a_d)$\@.

\subsection{Theory}
We start with the simple case where a full data set of $N$ points lies
within a $d$-dimensional unit hypercube $[0,1]^d$\@.  We further
assume that a random (or null) model would distribute these $N$ points
uniformly within the hypercube.  Additionally, we assume that the
hypercube is not periodic/toroidal.  We relax these simplifying
assumptions in subsequent sections.

We say that an axis-aligned, $d$-dimensional hypercube contained
within $[0,1]^d$ is a \ulem{box}\@.  We say that a set of $K \ge 2$ of
the $N$ points is a \ulem{$K$-boxed set} if there exists a box that
contains the $K$ points without containing any of the remaining $N-K$
points.  We define the \ulem{minimal box} for a $K$-boxed set as the
intersection of all boxes that contain the $K$ points.

A box of dimensions $(w_1, \ldots, w_d)$ or smaller with $K$ or more
points exists only if there is a $K$-boxed set that has a minimal box
with dimensions not greater than $(w_1, \ldots, w_d)$\@.  Thus, we aim
to compute the expected number of the latter.
We use as an integrand the probability density that a box will be a
minimal box for a set of $K$ points, and we integrate over all
applicable boxes.

Consider a box $[f_1, g_1] \times \ldots \times [f_d, g_d]$ with
dimensions $(l_1, \ldots, l_d) = (f_1 - g_1, \ldots, f_d - g_d)$ and
hypervolume
\begin{equation}
V = \prod_{i=1}^d l_i \eqnperiod
\end{equation}
It will be a (non-minimal) box indicating that a set of $K$ points is
a $K$-boxed set if exactly $K$ points fall within it.  Thus the
probability that it indicates a $K$-boxed set is
\begin{equation}
\left(\!\!\!\!\begin{array}{c}N\\K\end{array}\!\!\!\!\right) V^K
(1-V)^{N-K}
\label{eqn:binomial}
\eqnperiod
\end{equation}
However, it is our goal to calculate the probability density for a
\emph{minimal} box for $K$ points.  This box is minimal only if, for
every dimension index $i$, we have that $f_i$ is the minimum of the
$K$ points' $i$th coordinate values and $g_i$ is the maximum of these
coordinate values.  Thus, for each dimension, we must have one of the
$K$ points achieve the minimum and one of the remaining $K-1$ points
achieve the maximum.  The remaining $K-2$ coordinate values can be
anywhere in the range $(f_i, g_i)$.  (Note that the possibility of
having multiple points exactly achieve the minimum or maximum is an
event that has a probability density of zero and thus is safely
ignored.)  Thus, the $i$th dimension contributes a factor of
$K(K-1){l_i}^{K-2}$ to the probability density.  The probability
density for the event that the box is a minimal box for a $K$-boxed
set is
\begin{equation}
\Pr[K|l_1,\ldots,l_d] =
\left(\!\!\!\!\begin{array}{c}N\\K\end{array}\!\!\!\!\right) K^d (K-1)^d V^{K-2} (1-V)^{N-K}
\label{eqn:density}
\eqnperiod
\end{equation}

We note that a box with dimensions $(l_1, \ldots, l_d)$ can have its
minimum corner $(f_1, \ldots, f_d)$ anywhere in $[0, 1-l_1] \times
\ldots \times [0, 1-l_d]$\@.  When we restrict attention to minimal
boxes with dimensions not greater than $(w_1, \ldots, w_d)$ the
expected number of $K$-boxed sets is
\begin{eqnarray}
\lefteqn{E(K,w_1,\ldots,w_d) =} \nonumber \\
& & 
\int_0^{w_1} dl_1 \ldots \int_0^{w_d} dl_d
\left( \prod_{i=1}^d (1-l_i) \right)
\Pr[K|l_1,\ldots,l_d].
\label{eqn:Evalue}
\eqnperiod
\end{eqnarray}

Equation~\ref{eqn:Evalue} provides an upper bound to the probability
that random data will exhibit a clump of $K$ or more points in a box
of dimensions $(w_1, \ldots, w_d)$ or smaller, because such a box
exists only if there is a $K$-boxed set that has a minimal box with
dimensions not greater than $(w_1, \ldots, w_d)$\@.  Under many
circumstances, when $E \ll 1$ the probability of multiple $K$-boxed
sets is small and the bound is quite tight.

\subsection{Implementation}
We observe that
\begin{equation}
(1-V)^{N-K} = \sum_{m=0}^{N-K} \left(\!\!\!\!\begin{array}{c}N-K\\m
\end{array}\!\!\!\!\right)
(-V)^m \eqncomma
\end{equation}
and, to speed the integration using Maple~7, we seek a way to truncate
the sum at significantly fewer terms.  We observe that
$\left(\!\!\!\!\begin{array}{c}N-K\\m
\end{array}\!\!\!\!\right)V^m$ is roughly a bell curve as a function
of $m$.  The bell curve has mean and standard deviation given by
\begin{eqnarray}
\mu & = & \frac{(N-K)V}{1+V} \le NV \le K \\
\sigma & = & \sqrt{\frac{(N-K)V}{(1+V)^2}} \le \sqrt{NV} \le \sqrt{K}
\eqncomma
\end{eqnarray}
where we have assumed that $K \ge NV$, \ie, that a clump is under
consideration only when the number of points it contains is more than
the expected number.

Thus, we can include all terms of $(1-V)^{N-K}$ out to $z=10$ standard
deviations with $m^* = \max(\lceil \mu + z \sigma \rceil, 25)$ and the
precise approximation
\begin{equation}
(1-V)^{N-K} \approx \sum_{m=0}^{m^*} \left(\!\!\!\!\begin{array}{c}N-K\\m
\end{array}\!\!\!\!\right)
(-V)^m \eqnperiod
\end{equation}
We use this approximation in Equation~\ref{eqn:Evalue} and we evaluate
the integral with Maple~7\@.

\subsection{Variations}
\subsubsection{Non-Unit Hypercubes and Non-Uniform Random Distributions}
In many circumstances, including the case presented in this paper, the set of 
$N$ data points does not fall
uniformly into the unit hypercube under the random/null model.  For
instance, one does not expect the radial velocities of spheroid stars to be
uniformly distributed in any velocity range, and they are certainly not
limited to the range $[0,1]$. 
%points uniformly distributed within a sphere of radius $R$
%have $(r, \theta, \phi) \in [0,R] \times [0,\pi] \times [0, 2\pi]$
%with a single-point density of

We say that the density is \ulem{separable} if it can be written as
the product of functions, each of which depends on a single dimension
$\Pr(\vec{x}) \propto \prod_{i=1}^d h_i(x_i)$.  For instance, the
uniform spherical density,
\begin{equation}
\Pr(r,\theta,\phi) = \frac{1}{4\pi R^3/3} r^2 \sin(\theta)
\, dr \, d\theta \, d\phi \eqnperiod
\end{equation}
is separable.
%% because
%% \begin{equation}
%% \Pr(r, \theta, \phi) \propto f(r) g(\theta) h(\phi)
%% \end{equation}
%% for $f(r) = r^2 / (R^3/3)$, $g(\theta) = \sin(\theta)/2$, and $h(\phi) =
%% 1/2 \pi$\@.
When a probability density for a coordinate vector $\vec{x} \in [A_i,
B_i] \times \ldots \times [A_d, B_d]$ is separable, we can perform a
coordinate transformation to a coordinate vector $\vec{y}$ that is
uniformly distributed in the unit hypercube and which preserves
axis-aligned hypercubes:
\begin{equation}
y_i = \frac%
{\int_{A_i}^{x_i} h_i(x'_i) \, dx'_i}%
{\int_{A_i}^{B_i} h_i(x'_i) \, dx'_i}%
\eqnperiod
\end{equation}
The clumping question then becomes one of percentiles, \ie, what is
the probability that random data will show $K$ or more points in a
axis-aligned hypercube with dimensions that do not exceed $w_i$
percentiles $\times \ldots \times w_d$ percentiles.
In this paper, the radial velocities of the stars and the apparent
$g_0$ magnitudes were treated as separable dimensions.

\subsubsection{Periodic or Toroidal Boundary Conditions}
In many circumstances, it is desirable to allow $K$-boxed sets to wrap
around the boundaries in one or more dimensions.  In this paper, it
was desirable to allow a box to span the prime meridian in a longitude
coordinate.  In such situations the restriction that the minimum
corner $(f_1, \ldots, f_d)$ fall in $[0, 1-l_1] \times \ldots \times
[0, 1-l_d]$ is relaxed.  When the $i$th coordinate is periodic, $f_i$
can be any value in $[0,1]$\@.  Equation~\ref{eqn:Evalue} is thus
modified by removal of the $(1-l_i)$ factor for each periodic
dimension $i$\@.

\subsubsection{Poisson Distributions}
In many circumstances, it is desirable to have the random/null model
be Poisson; instead of requiring exactly $N$ points in the unit
hypercube, the number of points $n$ is selected using a Poisson
distribution with mean $N$:
\begin{equation}
\Pr[n | N] = e^{-N} \frac{N^n}{n!} \eqnperiod
\end{equation}
In our example, the Poisson distribution applies if there are
an average of 4149 BHB stars with spectra in the region $b>45^\circ$, whereas the 
uniformly random model would put exactly 4149 BHB stars with spectra in that same
region.

In the Poisson case, Expression~\ref{eqn:binomial} becomes
\begin{equation}
\frac{N^K}{K!} V^K e^{-N V} \eqncomma
\end{equation}
Equation~\ref{eqn:density} becomes
\begin{equation}
\Pr[K|l_1,\ldots,l_d] =
\frac{N^{K}}{K!} K^d (K-1)^d V^{K-2} e^{-N V} \eqncomma
\label{eqn:PoissonDensity}
\end{equation}
and Equation~\ref{eqn:Evalue} is unmodified, except that it
incorporates Equation~\ref{eqn:PoissonDensity} instead of
Equation~\ref{eqn:density}\@.

For integration, instead of truncating the series for $(1-V)^{N-K}$,
we truncate the series for
\begin{equation}
e^{-N V} = \sum_{m=0}^{\infty} \frac{(-NV)^m}{m!} \eqnperiod
\end{equation}
We observe that $(NV)^m/m!$ is roughly a bell curve as a function of
$m$.  The bell curve has mean and standard deviation given by
\begin{eqnarray}
\mu & = & NV \le K \\
\sigma & = & \sqrt{NV} \le \sqrt{K} \eqnperiod
\end{eqnarray}
Thus, we can include all terms of $\exp(-NV)$ out to $z$ standard
deviations with $m^* = \max(\lceil \mu + z \sigma \rceil, 25)$ and the
precise approximation
\begin{equation}
e^{-NV} \approx \sum_{m=0}^{m^*} \frac{(-NV)^m}{m!} \eqnperiod
\end{equation}
We use this approximation in Equation~\ref{eqn:PoissonDensity} and we
evaluate the integral of Equation~\ref{eqn:Evalue} with Maple~7\@.
We determined that for this paper the distinction between random
and Poisson distributions was unimportant.

\subsubsection{Non-Separability}
When the dimensions are not separable, the assumption that the joint
density function is the product of the individual dimension's density
functions does not hold.  However, if there is a set of individual
density functions for which the assumption is approximately true, we
define the \ulem{overdensity} to be the ratio of joint density
function to the value implied by multiplying the approximating density
functions together.  The calculations for the separable case make a
good approximation for the non-separable case if the percentile widths
$(w_1, \ldots, w_d)$ are adjusted by factors whose product is the
overdensity.  When the overdensity is greater than 1.0, a conservative 
approach is to multiply the smallest of the widths of the non-separable
dimensions by the overdensity.

\bsp

\label{lastpage}

\end{document}